\documentclass[12pt]{article}
\usepackage{amsmath,amsthm,amssymb,euscript,epsfig,youngtab}
\usepackage{array}
\usepackage{fancybox}
\usepackage[usenames]{color}
\usepackage{amsfonts,bm}
\usepackage{graphicx}

\newcommand{\be}{\begin{equation}}
\newcommand{\ee}{\end{equation}}
\newcommand{\bea}{\begin{eqnarray}\displaystyle}
\newcommand{\eea}{\end{eqnarray}}

\makeatletter
\@addtoreset{equation}{section}
\makeatother
\renewcommand{\theequation}{\thesection.\arabic{equation}}
\def\one{{\hbox{ 1\kern-.8mm l}}}
\def\zero{{\hbox{ 0\kern-1.5mm 0}}}

\begin{document}
\makeatletter
\@addtoreset{equation}{section}
\makeatother
\renewcommand{\theequation}{\thesection.\arabic{equation}}

\rightline{WITS-CTP-132}
\vspace{1.8truecm}

\vspace{15pt}


{\LARGE
\centerline{\bf  Heavy Operators in}
\centerline{\bf Superconformal Chern-Simons Theory}
}  

\vskip.5cm 

\thispagestyle{empty} \centerline{
    {\large \bf Robert de Mello Koch\footnote{{\tt robert@neo.phys.wits.ac.za}},}
   {\large \bf Rocky Kreyfelt\footnote{{\tt Rocky.Kreyfelt@students.wits.ac.za}}
    and Stephanie Smith\footnote{{\tt stephiiyy@gmail.com}}}
                                                       }
\vspace{.4cm}
\centerline{{\it National Institute for Theoretical Physics ,}}
\centerline{{\it School of Physics and Centre for Theoretical Physics }}
\centerline{{\it University of Witwatersrand, Wits, 2050, } }
\centerline{{\it South Africa } }

\vspace{1.4truecm}

\thispagestyle{empty}

\centerline{\bf ABSTRACT}

\vskip.4cm 

We study the anomalous dimensions for scalar operators in ABJM theory in the $SU(2)$ sector. 
The operators we consider have a classical dimension that grows as $N$ in the large $N$ limit.
Consequently, the large $N$ limit is not captured by summing planar diagrams - non-planar contributions have to be included.
We find that the mixing matrix at two-loop order is diagonalized using a double coset ansatz, reducing it to the Hamiltonian
of a set of decoupled oscillators.
The spectrum of anomalous dimensions, when interpreted in the dual gravity theory, shows that the energy of the fluctuations 
of the corresponding giant graviton is dependent on the size of the giant.
The first subleading corrections to the large $N$ limit are also considered.
These subleading corrections to the dilatation operator do not commute with the leading terms, indicating that integrability
may not survive beyond the large $N$ limit.

\setcounter{page}{0}
\setcounter{tocdepth}{2}

\newpage

\setcounter{footnote}{0}

\linespread{1.1}
\parskip 4pt

\section{Introduction}\label{intro}

Integrability has proven to be a powerful tool in analyzing the spectrum of anomalous dimensions in ${\cal N} = 4$ 
Super Yang-Mills theory in the planar limit\cite{Minahan:2002ve,Beisert:2010jr}. 
An interesting question is whether or not there are other large $N$ limits that are also integrable.
This question has been the focus of a number of recent 
studies\cite{Koch:2010gp,DeComarmond:2010ie,Carlson:2011hy,Koch:2011hb,deMelloKoch:2011ci,deMelloKoch:2012ck,deMelloKoch:2011vn,deMelloKoch:2012sv,Koch:2012sf,Koch:2013xaa,Koch:2013yaa,Lin:2014yaa}.
At this point there is evidence that suggests certain large $N$ limits, that are not captured by simply summing the planar diagrams, do enjoy integrability.

The studies described above have all focused on ${\cal N}=4$ super Yang-Mills theory.
In this article we extend existing studies by exploring a large $N$ but non-planar limit of the ABJM theory, which is an ${\cal N}=6$ superconformal Chern-Simons-matter 
theory with gauge group $U(N)\times U(N)$ on $R^{1,2}$ and Chern-Simons levels $k$ and $-k$.
Almost all of the results that have been obtained in the planar limit of ${\cal N} = 4$ super Yang-Mills theory hold in an appropriately modified form for the ABJM theory\cite{Klose:2010ki}.
Further, the technology needed to study operators with anomalous dimensions that grow as $N$ (called ``heavy operators'') has
been developed\cite{deMelloKoch:2012kv,Pasukonis:2013ts,Koch:2014yga}.
It is thus very natural to search for possible large $N$ but non-planar limits of ABJM theory that enjoy integrability.
This is the primary motivation for the study reported in this article.

We confine attention to the $SU(2)$ sector of theory and work at two loops.
In this case, relying on results of \cite{Koch:2014yga}, we are able to give a simple description, which employs restricted Schur polynomials. 
Concretely, \cite{Koch:2014yga} proved that a basis for the operators in this sector of the theory is provided by restricted
Schur polynomials in the adjoints (of one of the $U(N)$ factors) constructed out of the bifundamental scalars fields.
The delicate point, resolved in \cite{Koch:2014yga}, involves demonstrating that the finite $N$ constraints are correctly
accounted for.
Our polynomials employ two adjoints, called $\phi_{11}$ and $\phi_{12}$ below.
The number of $\phi_{11}$ fields is $n_{11}$ and the  number of $\phi_{12}$ fields is $n_{12}$.
As we show in section 2, the structure of the one loop dilatation operator for ABJM theory differs from that of ${\cal N}=4$ super 
Yang-Mills theory.
The operators we consider are labeled by Young diagrams with $O(1)$ rows or columns and a total of $O(N)$ boxes.
For these operators we can employ the displaced corners approximation of \cite{Koch:2011hb}.
This requires $n_{12}\gg n_{11}$.
In this approximation, the leading terms in the dilatation operator are diagonalized using a double coset ansatz\cite{deMelloKoch:2012ck} and the results of spring field theory\cite{deMelloKoch:2011ci}.
The dilatation operator reduces to a set of decoupled oscillators.
There are subleading terms of size ${n_{11}\over n_{12}}$ relative to the leading contribution, which represent corrections to 
the large $N$ limit.
These subleading terms are not diagonalized by the ansatz of \cite{deMelloKoch:2012ck}, so that a careful treatment
of these terms would indicate whether the large $N$ but non-planar integrability is a property only of the large $N$ limit.
Our study shows that these subleading terms do not commute with the leading order, so that they are not diagonalized
by the ansatz of \cite{deMelloKoch:2012ck}.
Although this does not prove that the system is not integrable, it does suggest that the integrability we have found
is only a property of the large $N$ limit.
Given similar results obtained in the planar limit of the theory\cite{Kristjansen:2008ib,Caputa:2009ug}, this is not surprising.
In last section we summarize our results and point out some interesting directions in which this study can be extended.

There are a number of further works related to our study, with relevant background.
In particular, \cite{Berenstein:2008dc} lays the foundation for the description of membranes in ABJM using a group
theoretic perspective.
See also \cite{Berenstein:2006qk,de Mello Koch:2007uu} for background from the ${\cal N}=4$ super Yang-Mills 
theory which is relevant for our study. 

\section{$SU(2)$ Dilatation Operator in Adjoint Variables}\label{advars}

We are studying an ${\cal N}=6$ Chern-Simons gauge theory with $U(N)\times U(N)$ gauge group.
The generalized restricted Schur polynomials, introduced and studied in \cite{Pasukonis:2013ts} provide a  basis for the  local operators of any quiver gauge theory with gauge group built from unitary group factors.
In constructing our local operators we will use scalar fields $A_1,A_2$ both transforming in the $(N,\bar{N})$ of $U(N)\times U(N)$, as well
as $B_1^\dagger,B_2^\dagger$ which transform in the $(\bar{N},N)$.
Given these transformation properties, it is clear that the fields 
$$
\phi_{11}{}^a_b =A_1{}^a_\alpha B_1^\dagger{}^\alpha_b \, ,
\qquad
\phi_{12}{}^a_b =A_1{}^a_\alpha B_2^\dagger{}^\alpha_b \, ,
$$
$$
\phi_{21}{}^a_b =A_2{}^a_\alpha B_1^\dagger{}^\alpha_b \, ,
\qquad
\phi_{22}{}^a_b =A_2{}^a_\alpha B_2^\dagger{}^\alpha_b\, .
$$
transform in the adjoint of the first $U(N)$ and as a singlet of the second.
In general, the description of the theory in terms of these adjoint fields does not correctly capture the finite $N$ physics.
Indeed, as explained in \cite{Koch:2014yga}, the constraints on local operators at finite $N$ arising from the fact that the adjoints are $N\times N$ matrices is a subset of the full set of constraints, arising 
because both $A_I$ and $B_I^\dagger$ are $N\times N$ matrices.
However, if we restrict to the so called $SU(2)$ sector in which only $\phi_{11}$ and $\phi_{12}$ are used, the finite $N$ constraints resulting from the description employing adjoint scalars $\phi_{11}$ and $\phi_{12}$
agree with the constraints obtained from the original variables.
The description employing adjoints has the advantage that the restricted Schur polynomials of \cite{deMelloKoch:2012kv} provides a suitable basis, and the technology to work with these operators is well developed (see for example \cite{Koch:2011hb}).
The restricted Schur polynomials we use are
\bea
   \chi_{R,\{ r\},\alpha\beta}(\phi_{11},\phi_{12})
    =\frac{1}{n_{11}!n_{12}!}\sum_{\sigma \in S_{m_1+m_2}}{\rm Tr}_{\{r\},\alpha\beta}(\Gamma_R(\sigma)) Tr\Big(\sigma (\phi_{11})^{\otimes n_{11}}(\phi_{12})^{\otimes n_{12}}\Big)
   \label{ourOps}
\eea
where we are considering an operator constructed using $n_{11}$ $\phi_{11}$ fields and $n_{12}$ $\phi_{12}$ fields.
$\{ r\}$ denotes an irreducible representation of $S_{n_{11}}\times S_{n_{12}}\subset S_{n_{11}+n_{12}}$.
It is useful to think of $\{r\}$ as a pair of Young diagrams, one with $n_{11}$ boxes and one with $n_{12}$ boxes.
The irreducible representation $\{r\}$ may appear more than once upon restricting the 
representation $R$ of $S_{n_{11}+n_{12}}$ to the $S_{n_{11}}\times S_{n_{12}}$ subgroup.
The multiplicity labels $\alpha,\beta$ distinguish between these different copies.
The trace ${\rm Tr}_{\{r\},\alpha\beta}(\Gamma_R(\sigma))$ is an instruction to trace only over the $\{r\}$ subspace within the carrier space of $R$.
Further, row indices are traced over the $\alpha$ copy of $\{r\}$ while the column indices are traced over the $\beta$ copy.
To implement the restricted trace we introduce intertwining operators $P_{R,\{r\},\alpha\beta}$ defined so that
\bea
  {\rm Tr}_R\Big(P_{R,\{r\},\alpha\beta}\Gamma_R(\sigma)\Big)={\rm Tr}_{\{r\},\alpha\beta}\Big(\Gamma_R(\sigma)\Big)
\eea
where the trace on the LHS now runs over the full carrier space of $R$.
Our conventions for the action of the symmetric group in the space $V^{\otimes n_{11}+n_{12}}$ on which the multilinear operators $(\phi_{11})^{\otimes n_{11}}(\phi_{12})^{\otimes n_{12}}$ act 
are as follows
\bea
    (\sigma)^I_J =\delta^{i_1}_{j_{\sigma(1)}}\cdots\delta^{i_{n_{11}+n_{12}}}_{j_{\sigma(n_{11}+n_{12})}}
\eea
The two point function of these operators is\cite{deMelloKoch:2012kv}
\bea
\langle    \chi_{R,\{ r\},\alpha\beta}(\phi_{11},\phi_{12})   \chi_{S,\{ s\},\gamma\delta}(\phi_{11},\phi_{12})^\dagger  \rangle
=\delta_{RS}\delta_{r_{11}s_{11}}\delta_{r_{12}s_{12}}\delta_{\alpha\gamma}\delta_{\beta\delta}{f_R^2{\rm hooks}_R\over{\rm hooks}_{r_{11}}{\rm hooks}_{r_{12}}}
\label{twopoint}
\eea
We will need this result below.

The dilatation operator, acting in this $SU(2)$ sector, is given by\cite{Kristjansen:2008ib}
\bea
  D=-\left( {4\pi\over k}\right)^2 :{\rm Tr}
     \Big[ \left(B_2^\dagger A_1 B_1^\dagger - B_1^\dagger A_1 B_2^\dagger\right)\left(
            {\partial\over\partial B_2^\dagger}{\partial\over\partial A_1}{\partial\over\partial B_1^\dagger} 
         -  {\partial\over\partial B_1^\dagger}{\partial\over\partial A_1}{\partial\over\partial B_2^\dagger}\right)\Big]:
\eea
A straightforward application of the chain rule allows us to rewrite this in terms of adjoint fields as\footnote{For the ABJ theory
with gauge group $U(N)\times U(M)$, assuming $M>N$, the only change in this formula is that the factor of $N$ in the third last 
line of (\ref{newdil}) would be replaced by an $M$. The $\phi_{ij}$ would continue to be $N\times N$ matrices. If $N<M$ our
description changes as we would need to form $U(M)$ adjoints.}
\bea
&& :{\rm Tr} \Big[ \left(B_2^\dagger A_1 B_1^\dagger - B_1^\dagger A_1 B_2^\dagger\right)\left(
            {\partial\over\partial B_2^\dagger}{\partial\over\partial A_1}{\partial\over\partial B_1^\dagger} 
         -  {\partial\over\partial B_1^\dagger}{\partial\over\partial A_1}{\partial\over\partial B_2^\dagger}\right)\Big]:\cr
&&= :{\rm Tr}
     \Big[\left(\phi_{12}\phi_{11}-\phi_{11}\phi_{12}\right)\left(
            {\partial\over\partial\phi_{12}}\phi_{1j} {\partial\over\partial\phi_{1j}}{\partial\over\partial\phi_{11}} 
         -  {\partial\over\partial\phi_{11}}\phi_{1j}{\partial\over\partial\phi_{1j}}{\partial\over\partial\phi_{12}}\right)\Big]:\cr
&&+N : {\rm Tr}     \Big[\left(\phi_{12}\phi_{11}-\phi_{11}\phi_{12}\right)\left(
            {\partial\over\partial\phi_{12}}{\partial\over\partial\phi_{11}} 
         -  {\partial\over\partial\phi_{11}}{\partial\over\partial\phi_{12}}\right)\Big]:\cr
 &&+:{\rm Tr}\left[(\phi_{12}\phi_{11}-\phi_{11}\phi_{12}){\partial\over\partial\phi_{12}}\right]{\rm Tr}\left[{\partial\over\partial \phi_{11}}\right]:\cr
&&-:{\rm Tr}\left[(\phi_{12}\phi_{11}-\phi_{11}\phi_{12}){\partial\over\partial\phi_{11}}\right]
{\rm Tr}\left[{\partial\over\partial \phi_{12}}\right]:
\label{newdil}
\eea
We now turn to the problem of evaluating the action of the dilatation generator on the operators (\ref{ourOps}).
The evaluation uses the technology developed in \cite{DeComarmond:2010ie,Koch:2011hb}.
The matrix derivatives are straight forward to evaluate; in manipulating the resulting expressions the identity
$$
   {\rm Tr}(\rho\cdot\alpha\cdot\beta \phi^{\otimes n})=\prod_{A=1}^n\phi^{l_{\beta^{-1}(A)}}_{l_{(\alpha\rho)(A)}}
$$
is extremely useful.
To express the result of the action of $D$ as a linear combination of restricted Schur polynomials, a key ingredient is the identity 
$$
{\rm Tr}(\tau \phi_{11}^{\otimes n_{11}}\phi_{12}^{\otimes n_{12}})
=\sum_{R,\{r\},\alpha\beta}
{d_R n_{11}!n_{12}!\over d_{r_{11}}d_{r_{12}}\, n!}
\chi_{R,\{ r\},\alpha\beta }(\tau) \chi_{R,\{ r\},\beta\alpha}
$$
where the sum over $R$ runs over all irreducible representations of $S_{n_{11}+n_{12}}$ and $\{ r\}$ is summed over all 
irreducible representations of $S_{n_{11}} \times S_{n_{12}}$.
This identity is derived in \cite{Bhattacharyya:2008xy} in the context of $U(N)$ gauge theory and it applies without change to our description in terms of adjoints.
We are interested in operators with a bare dimension of order $N$. 
We achieve this large dimension by taking $n_{12}$ order $N$ and $n_{11}$ order $\sqrt{N}$.
For these operator, not all terms in (\ref{newdil}) have the same size at large $N$.
The sizes of the different terms follow by noting that differentiating with respect to $\phi_{12}$ produces order $N$ terms
while differentiating with respect to $\phi_{11}$ produces order $\sqrt{N}$ terms. Consequently, in the first term of
(\ref{newdil}) the terms with $j=2$ dominate; the terms with $j=1$ are supressed by a relative factor of $\sqrt{N}$.
Apart from the leading term, we will also study this first subleading contribution in this work.
The second term in (\ref{newdil}) also contributes at the leading order. 
The third and fourth terms in (\ref{newdil}) are subleading, supressed by ${1\over N}$ and will consequently not be considered further in our study.
It would not be consistent to evaluate these terms without also including the ${1\over N}$ correction to the leading terms.
Finally, it is useful to express our result in terms of operators normalized so that
\begin{equation}
\langle\hat{O}_{R,\{ r\},\alpha\beta} \hat{O}_{S,\{ s\},\gamma\delta}^\dagger \rangle=f_R \delta_{RS}\delta_{r_{11}s_{11}}\delta_{r_{12}s_{12}}\delta_{\alpha\gamma}\delta_{\beta\delta}
\label{norm}
\end{equation}
Clearly then
\bea
  \hat{O}_{R,\{ r\},\alpha\beta} (\phi_{11},\phi_{12})=\sqrt{f_R {\rm hooks}_R\over {\rm hooks}_{r_{11}}{\rm hooks}_{r_{12}}} \chi_{R,\{ r\},\alpha\beta}(\phi_{11},\phi_{12})
\eea
The normalization in (\ref{norm}) has been chosen so that the leading contribution to the dilatation operator most
closely resembles the result obtained in \cite{DeComarmond:2010ie} for ${\cal N}=4$ super Yang-Mills theory.
Note that operators labeled by Young diagrams $R$ with different shapes, are not normalized in the same way.
Clearly, from (\ref{norm}) it follows that the ratio of their normalizations is given by the ratios of the factors of the boxes 
that do not agree between the two labels. 
For operators with a dimension of order $N$ and number of rows (or columns) of order 1, this ratio is always equal to 1 plus
${1\over N}$ corrections. 
Putting these ingredients together, we find
\bea
   D\hat{O}_{R,\{ r\},\alpha\beta} &&=\sum_{S,\{ s\}\gamma\delta}
\sqrt{f_S{\rm hooks}_S{\rm hooks}_{r_{11}}{\rm hooks}_{r_{12}}\over f_R{\rm hooks}_R{\rm hooks}_{s_{11}}{\rm hooks}_{s_{12}}}
M_{R,\{r\},\alpha\beta;S,\{s\},\gamma\delta}\hat{O}_{S,\{s\},\gamma\delta}\cr
&&\equiv \sum_{S,\{ s\}\gamma\delta}\, D_{R,\{ r\},\alpha\beta\, ;\, S,\{s\},\gamma\delta}\hat{O}_{S,\{s\},\gamma\delta}
\label{dilop1}\eea
where
\bea
&   M_{R,\{r\},\alpha\beta;S,\{s\},\gamma\delta}=-\left( {4\pi\over k}\right)^2
\sum_{R'}{c_{RR'}d_S n_{11} n_{12}\over d_{s_{11}}d_{s_{12}}d_{R'}(n_{11}+n_{12})}\cr
&\times\Bigg[ (n_{12}-1){\rm Tr}_{R\oplus S} \bigg[I_{S'R'}(1,n_{11}+2)\big[ (1,n_{11}+1),
P_{R,\{r\}\alpha\beta}\big]I_{R'S'}\big[(1,n_{11}+1),P_{S,\{s\}\gamma\delta}\big]\bigg]\cr
&+(n_{11}-1) {\rm Tr} \bigg[I_{S'R'}(1,2)\big[ (1,n_{11}+1),P_{R,\{r\}\alpha\beta}\big]I_{R'S'}
\big[(1,n_{11}+1),P_{S,\{s\}\gamma\delta}\big]\bigg]\cr
&+ N {\rm Tr} \bigg[I_{S'R'}\big[ (1,n_{11}+1),P_{R,\{r\}\alpha\beta}\big]I_{R'S'}
\big[(1,n_{11}+1),P_{S,\{s\}\gamma\delta}\big]\bigg]\cr
&+  {\rm Tr} \bigg[I_{S'R'}\left(
P_{R,\{r\}\alpha\beta}-(1,n_{11}+1)P_{R,\{r\}\alpha\beta}(1,n_{11}+1)\right)
I_{R'S'}\big[(1,n_{11}+1),P_{S,\{s\}\gamma\delta}\big]\bigg]
\Bigg]\cr
\label{dilop2}
\eea
To obtain this result, the sum over the symmetric group appearing in (\ref{ourOps}) is evaluated using the fundamental 
orthogonality theorem of group representation theory.
The sum that appears after the derivatives act is a sum over $S_{n_{11}+n_{12}-1}\subset S_{n_{11}+n_{12}}$, so that the sum is non-zero as long as one of the representations
suduced by $R$ upon restricting to $S_{n_{11}+n_{12}-1}$ agrees with one of the representations subduced by $S$
under the same restriction.
The sum then produces the maps $I_{S'R'}$ and $I_{R'S'}$ which map between subspaces of the carrier spaces of $R$ and $S$. 
We have used cycle notation for elements of the symmetric group.
To completely spell out our notation, note that each element of the symmetric group is in the representation inherited from
the subspace it acts in.
Thus, for example,
\bea 
{\rm Tr}_{R\oplus S} \bigg[I_{S'R'}(1,n_{11}+2)I_{R'S'}(1,n_{11}+1)\bigg]
={\rm Tr}_{R\oplus S} \bigg[I_{S'R'}\Gamma^{R}\left((1,n_{11}+2)\right) I_{R'S'}
\Gamma^{S}\left((1,n_{11}+1)\right)\bigg]\nonumber
\eea
where $\Gamma^{S}\left(\sigma\right)$ is the matrix representing $\sigma$ in irreducible representation $S$.

The formulas (\ref{dilop1}) and (\ref{dilop2}) are the key results of this section.
These are exact in the sense that we have not used any simplifications of the large $N$ limit to obtain this result.
We now consider the eigenproblem of $D$ which, as we explain in the next section, can be solved in a specific limit, after exploiting simplifications of large $N$.
At large $N$ the last line in (\ref{dilop2}) is subleading and will therefore be dropped in what follows\footnote{The last line in (\ref{dilop2}) corresponds to the
third and fourth terms in (\ref{newdil})}.

\section{Displaced Corners Approximation}

It is perhaps useful to begin with a discussion of some of the intricacies inherint in the problem of diagonalizing (\ref{dilop1}).
The key difficulty in constructing the restricted Schur polynomials (\ref{ourOps}) is in the construction of the intertwining operators $P_{R,\{r\},\alpha\beta}$. 
To compute the two point function (\ref{twopoint}), after summing over the free field Wick contractions, we simply need to take a product of two of these intertwining operators and then compute their trace, which is a relativly simple computation.
Indeed, the result depends only on the dimensions of the representations $R$ and $\{ r\}$ which appear.
The expression in (\ref{dilop1}) involves computing commutators of the intertwining operators with symmetric group elements and then tracing over a product of these commutators.
This is a much more sophisticated operation for which the explicit form of $P_{R,\{r\},\alpha\beta}$ is required.
Fortunately there is a limit in which we can construct $P_{R,\{r\},\alpha\beta}$ in a straight forward way: this is the displaced corners limit of \cite{Koch:2011hb} (see also \cite{Carlson:2011hy}).
The idea is simply that for the vast majority of restricted Schur polynomials 
$\chi_{R,\{ r\},\alpha\beta}(\phi_{11},\phi_{12})$ that can be written down, the distance 
between the last box in each row of $R$ is order $N$.
Here by the distance between boxes $a$ and $b$ we mean the smallest number of boxes that one needs to pass through when moving, in the Young diagram, from box $a$ to box $b$.
When the distance between the last box in the different rows of $R$ is order $N$, the action of the symmetric group simplifies
dramatically, which greatly simplifies the construction of $P_{R,\{r\},\alpha\beta}$.
To guarantee this simplification it is necessary to assume in addition that $n_{12}\gg n_{11}$; for
further discussion and all the details see \cite{Koch:2011hb}.
In this article we accomplish $n_{12}\gg n_{11}$ by scaling $n_{12}$ as $N$ and $n_{11}$ as $\sqrt{N}$ as we take $N\to\infty$.
Our results would seem to hold with $n_{11}$ scaled as $N^{\alpha}$ with $\alpha<1$, but due to the formidable technical
computations needed, we have not managed to explore this important point in detail.
For a Young diagram $R$ with $p$ rows, the maps $I_{S'R'}$ and $I_{R'S'}$ can be identified with elements of $u(p)$.
The action of the symmetric group elements appearing in (\ref{dilop1}), on these maps, is easy to evaluate.
The intertwining operators themselves take a factorized form
\bea
    P_{R,\{r\},\alpha\beta}=p_{r_{11}\alpha\beta}{\bf 1}_{r_{12}}
\eea
where $p_{r_{11}\alpha\beta}$ projects onto $S_{n_{11}}$ irreducible representation 
$r_{11}$ and ${\bf 1}_{r_{12}}$ projects onto $S_{n_{12}}$ irreducible representation $r_{12}$.
The concrete construction of these intertwining operators, together with detailed examples, is given in \cite{Koch:2011hb}.

Since we have to take $n_{12}\gg n_{11}$ we know that the terms in (\ref{newdil}) with $j=2$ will dominate.
This is indeed the case: in (\ref{dilop2}) the terms with coefficient $n_{12}-1$ come from the $j=2$ term of (\ref{newdil}) while the terms with coefficient $n_{11}-1$ come from $j=1$.
In this section we will restrict our attention to large $N$, which implies that we should keep only the leading order in 
${n_{11}\over n_{12}}$.
This amounts to keeping only the terms in (\ref{dilop2}) that have coefficient $n_{12}-1$ or coefficient $N$
\bea
   D^{(0)}\hat{O}_{R,\{ r\},\alpha\beta} &&=\sum_{S,\{ s\}\gamma\delta}
\sqrt{f_S{\rm hooks}_S{\rm hooks}_{r_{11}}{\rm hooks}_{r_{12}}\over f_R{\rm hooks}_R{\rm hooks}_{s_{11}}{\rm hooks}_{s_{12}}}
M^{(0)}_{R,\{r\},\alpha\beta;S,\{s\},\gamma\delta}\hat{O}_{S,\{s\},\gamma\delta}\cr
&&\equiv \sum_{S,\{ s\}\gamma\delta}\, D^{(0)}_{R,\{ r\},\alpha\beta\, ;\, S,\{s\},\gamma\delta}\hat{O}_{S,\{s\},\gamma\delta}
\label{Ldilop1}
\eea
where
\bea
&   M^{(0)}_{R,\{r\},\alpha\beta;S,\{s\},\gamma\delta}=-\left( {4\pi\over k}\right)^2\sum_{R'}{c_{RR'}d_S n_{11} n_{12}\over d_{s_{11}}d_{s_{12}}d_{R'}(n_{11}+n_{12})}\cr
&\times\Bigg[ (n_{12}-1){\rm Tr} \bigg[I_{S'R'}(1,n_{11}+2)\big[ (1,n_{11}+1),
P_{R,\{r\}\alpha\beta}\big]I_{R'S'}\big[(1,n_{11}+1),P_{S,\{s\}\gamma\delta}\big]\bigg]\cr
&+ N {\rm Tr} \bigg[I_{S'R'}\big[ (1,n_{11}+1),P_{R,\{r\}\alpha\beta}\big]I_{R'S'}
\big[(1,n_{11}+1),P_{S,\{s\}\gamma\delta}\big]\bigg]\Bigg]
\label{Ldilop2}
\eea
We will return to the term with coefficient $n_{11}-1$ in the next section.
In the displaced corners approximation, using the simplifcations just outlined, we obtain
\bea
D^{(0)}_{R,\{ r\},\alpha\beta\, ;\, S,\{s\},\gamma\delta}=
-\left( {4\pi\over k}\right)^2
\sqrt{f_S\over f_R}\sum_{R'}{c_{RR'}\over (n_{11}-1)!}(N+r_{12i}) \sqrt{{\rm hooks}_{r_{11}} {\rm hooks}_{s_{11}}}\times\cr
\Bigg[
{\rm Tr}(E^{(1)}_{kk}p_{r_{11}\alpha\beta}E^{(1)}_{ii}p_{s_{11}\gamma\delta})\delta_{r'_{12,i};s'_{12,k}}+
{\rm Tr}(E^{(1)}_{ii}p_{r_{11}\alpha\beta}E^{(1)}_{kk}p_{s_{11}\gamma\delta})\delta_{r'_{12,i};s'_{12,k}}\cr
-\Big({\rm Tr}(E^{(1)}_{kk}p_{r_{11}\alpha\delta})\delta_{\beta\gamma}+
 {\rm Tr}(E^{(1)}_{kk}p_{r_{11}\gamma\beta})\delta_{\alpha\delta}\Big)\delta_{R;S}\delta_{r_{11};s_{11}}\delta_{r_{12};s_{12}}
\Bigg]
\eea 
In this last formula, $r_{12i}$ is the length of row $i$ of Young diagram $r_{12}$, $R'$ is obtained from $R$ by dropping the last
box in row $i$ and $S'$ is obtained from $S$ by dropping the last box in row $k$.
$D^{(0)}_{R,\{ r\},\alpha\beta\, ;\, S,\{s\},\gamma\delta}$ is diagonalized by the double coset ansatz \cite{deMelloKoch:2012ck}.

To motivate what follows, recall that the label $\{ r\}=\{r_{11},r_{12}\}$ and that $r_{12}$ can be obtained by removing
a total of $n_{11}$ boxes from $R$.
Denote the number of rows in $R$ by $p$.
If we remove $a_1$ boxes from the first row, $a_2$ from the second and so on up to $a_p$ from row $p$, then the vector
$\vec{n}_{11}=(a_1,a_2,...,a_p)$ plays an important role: in the displaced corners approximation, operators with different
$\vec{n}_{11}$ do not mix at one loop \cite{Koch:2011hb}.
Of course, we have $a_1+a_2+\cdots+a_p = n_{11}$.
The vector $\vec{n}_{11}$ can be used to define a group $H$ which is a product of symmetric groups
\begin{equation}
   H=S_{a_1}\times S_{a_2}\times \cdots \times S_{a_p}
\end{equation}
According to the double coset ansatz\cite{deMelloKoch:2012ck}, each eigenfunction of the dilatation operator is in 
one-to-one correspondence with an element of the double coset $H\setminus S_{n_{11}}/ H$.
These double coset elements can also be put into correspondence with graphs whose edges are oriented and hence with
open strings states that obey the Gauss Law, providing a convincing connection with the dual D-brane plus open string
excited states; for background see \cite{Balasubramanian:2004nb,deMelloKoch:2012ck}.
The graph has a total of $p$ nodes and there are $n_{11}$ oriented edges stretching between the nodes.
For this reason we will refer to these operators as Gauss graph operators and to the associated oriented graphs as Gauss graphs.
The Gauss graph operators are\cite{deMelloKoch:2012ck}
\begin{equation}
O_{R,r_{12}}(\sigma)={|H|\over\sqrt{n_{11}!}}\sum_{j,k}\sum_{r_{11}\vdash n_{11}}\sum_{\mu_1,\mu_2}\sqrt{d_{r_{11}}}
\Gamma^{(r_{11})}_{jk}(\sigma)B^{r_{11}\to 1_H}_{j\mu_1}B^{r_{11}\to 1_H}_{k\mu_2}
\hat{O}_{R,\{ r\},\mu_1\mu_2}
\label{GGops}
\end{equation}
where $\sigma\in H\setminus S_{n_{11}}/ H$, $\Gamma^{(r_{11})}_{jk}(\sigma)$ is the matrix representing $\sigma$ in the
irreducible representation $r_{11}$ of $S_{n_{11}}$ and the branching coefficients $B^{r_{11}\to 1_H}_{j\mu_1}$ resolve
the projector from irreducible representation $r_{11}$ of $S_{n_{11}}$ to the trivial representation of $H$
\begin{equation}
{1\over |H|}\sum_{\gamma\in H}\Gamma^{(r_{11})}_{jk}(\sigma)=\sum_\mu B^{r_{11}\to 1_H}_{j\mu}B^{r_{11}\to 1_H}_{k\mu}
\end{equation}
Note that these operators are not normalized.
We have computed the norm of these operators in the Appendix.

The action of the dilatation operator is most easily written in terms of parameters read from the Gauss graphs.
Following \cite{deMelloKoch:2011uq}, a useful combinatoric description of a Gauss graph is obtained by dividing 
each string into two halves with a label for each half.
Using the orientation of the string, label both the outgoing and the ingoing string endpoints with an integer $1,2,\cdots,n_{11}$. 
A permutation is then determined by how the halves are joined and conversely, given a permutation, we can reconstruct the graph.
A graph is not associated to a unique permutation because the strings leaving the $i$'th node are indistinguishable, 
and the strings arriving at the $i$'th node are indistinguishable.
As a result, graphs are in one-to-one correspondence with elements of the double coset $H\setminus S_{n_{11}}/ H$.
Divide the integers $1,2,\cdots,n_{11}$ into $p$ sets, ${\cal S}_i$ $i=1,2,\cdots,p$ such that the symmetric group that is the
i$^{th}$ factor in $H$ permutes the elements of ${\cal S}_i$.
In the graph corresponding to $\sigma$, the number of oriented edges stretching from node $i$ to node $j$ is
\begin{equation}
   n_{ij}^+(\sigma)=\sum_{k\in{\cal S}_i}\sum_{l\in{\cal S}_j}\delta (\sigma (k),l)
\end{equation}
The number of strings stretching in the opposite direction, between the same two nodes, is
\begin{equation}
   n_{ij}^-(\sigma)=\sum_{k\in{\cal S}_i}\sum_{l\in{\cal S}_j}\delta (\sigma (l),k)
\end{equation}
The total number of strings stretching between the two nodes is $n_{ij}(\sigma)=n_{ij}^+(\sigma)+n_{ij}^-(\sigma)$.

The action of the dilatation operator is naturally written in terms of an operator $\Delta_{ij}$ defined as follows:
$\Delta_{ij}$ is a sum of three terms
\begin{equation}
  \Delta_{ij}=\Delta^+_{ij}+\Delta^0_{ij}+\Delta^-_{ij}
\end{equation}
To define the action of each of the above terms, we need to introduce two new Young diagrams,
$(r_{12})_{ij}^{\pm}$: $(r_{12})_{ij}^+$ is the Young
diagram obtained from $r_{12}$ by removing the last box from row $j$ and adding it to the end of 
row $i$, while $(r_{12})_{ij}^-$ is the Young
diagram obtained from $r_{12}$ by removing the last box from row $i$ and adding to the end of row $j$. 
$R^\pm_{ij}$ are defined in the same way.
The actions we need to define are\footnote{The $O(1)$ corrections added to $N$ in the expressions which follow
must be retained. After cancelations, these terms give the leading contribution.}
\begin{eqnarray}
&&\Delta^0_{ij}O_{R,r_{12}}(\sigma)=-(2N+r_{12i}+r_{12j}-3)O_{R,r_{12}}(\sigma)\cr
&&\Delta^+_{ij}O_{R,r_{12}}(\sigma)=\sqrt{(N+r_{12i}-1)(N+r_{12j}-1)}O_{R^+_{ij},(r_{12})^+_{ij}}(\sigma)\cr
&&\Delta^-_{ij}O_{R,r_{12}}(\sigma)=\sqrt{(N+r_{12i})(N+r_{12j}+2)}O_{R^-_{ij},(r_{12})^-_{ij}}(\sigma)
\end{eqnarray}
Recall that $r_{12k}$ is the number of boxes in row $k$ of Young diagram $r_{12}$.
A computation very similar to that of \cite{deMelloKoch:2012ck} now shows
\begin{equation}
D^{(0)}O_{R,r_{12}}(\sigma_1)=-\left( {4\pi\over k}\right)^2\sum_{\gamma_1,\gamma_2\in H}
\delta (\gamma\sigma_1\gamma^{-1}\sigma_2^{-1})\sum_{i<j}(N+r_{12,i})n_{ij}(\sigma_1)\Delta_{ij}O_{R,r_{12}}(\sigma_2)
\label{LeadDil}
\end{equation}
In the large $N$ limit we can introduce continuous variables $x_i$ defined by
\begin{equation}
x_i ={r_{12,i}-r_{12,p}\over \sqrt{N+r_{12,p}}}
\end{equation}
In terms of this continuous variable, the leading contribution to the action of the dilatation operator (\ref{LeadDil}) becomes
\bea
D^{(0)}O_{R,r_{12}}(\sigma_1)&=&-\left( {4\pi\over k}\right)^2\sum_{\gamma_1,\gamma_2\in H}
\delta (\gamma\sigma_1\gamma^{-1}\sigma_2^{-1})\cr
&\times&\sum_{i<j}(N+r_{12,i})n_{ij}(\sigma_1)
\left(\left({d\over dx_i}-{d\over dx_j}\right)^2-{(x_i-x_j)^2\over 4}\right)
O_{R,r_{12}}(\sigma_2)\cr
&&\label{fdil}
\eea
After diagonalizing $n_{ij}(\sigma)$ this is a sum of decoupled oscillators, which is an integrable system.

\section{Subleading term}\label{LstRSP}

In this section we will consider the subleading correction contained in
\bea
   D^{(1)}\hat{O}_{R,\{ r\},\alpha\beta} &&=\sum_{S,\{ s\}\gamma\delta}
\sqrt{f_S{\rm hooks}_S{\rm hooks}_{r_{11}}{\rm hooks}_{r_{12}}\over f_R{\rm hooks}_R{\rm hooks}_{s_{11}}{\rm hooks}_{s_{12}}}
M^{(1)}_{R,\{r\},\alpha\beta;S,\{s\},\gamma\delta}\hat{O}_{S,\{s\},\gamma\delta}\cr
&&\equiv \sum_{S,\{ s\}\gamma\delta}\, D^{(1)}_{R,\{ r\},\alpha\beta\, ;\, S,\{s\},\gamma\delta}\hat{O}_{S,\{s\},\gamma\delta}
\label{Ldilop3}
\eea
where
\bea
&   M^{(0)}_{R,\{r\},\alpha\beta;S,\{s\},\gamma\delta}=-\left( {4\pi\over k}\right)^2\sum_{R'}{c_{RR'}d_S n_{11} n_{12}\over d_{s_{11}}d_{s_{12}}d_{R'}(n_{11}+n_{12})}\cr
&\times (n_{11}-1) {\rm Tr} \bigg[I_{S'R'}(1,2)\big[ (1,n_{11}+1),P_{R,\{r\}\alpha\beta}\big]
I_{R'S'}\big[(1,n_{11}+1),P_{S,\{s\}\gamma\delta}\big]\bigg]
\label{Ldilop4}
\eea
These terms correspond to the terms with $j=1$ in (\ref{newdil}).
Evaluating the above trace in the displaced corners approximation, we find
\bea
D^{(1)}_{R,\{ r\},\alpha\beta\, ;\, S,\{s\},\gamma\delta}=
-\left( {4\pi\over k}\right)^2
\sqrt{f_S\over f_R}\sum_{R'}{c_{RR'}\over (n_{11}-2)!} \sqrt{{\rm hooks}_{r_{11}} {\rm hooks}_{s_{11}}}\times\cr
\Bigg[\sqrt{r_{12b}\over r_{12k}}
{\rm Tr}(E^{(1)}_{kk}E^{(2)}_{bi}p_{r_{11}\alpha\beta}E^{(1)}_{ib}p_{s_{11}\gamma\delta})\delta_{r'_{12,b};s'_{12,k}}+
{\rm Tr}(E^{(1)}_{id}E^{(2)}_{id}p_{r_{11}\alpha\beta}E^{(1)}_{kk}p_{s_{11}\gamma\delta})\delta_{r'_{12,i};s'_{12,k}}\cr
-\Big({\rm Tr}(E^{(1)}_{kb}E^{(2)}_{bk}p_{r_{11}\alpha\delta})
\delta_{ik}\delta_{r_{11}s_{11}}\delta_{\beta\gamma}\delta_{R;S}+
\sqrt{r_{12k}\over r_{12i}} {\rm Tr}(E^{(2)}_{ki}p_{r_{11}\gamma\beta}E^{(1)}_{ik}p_{s_{11}\gamma\delta})\Big)
\delta_{r_{12};s_{12}}
\Bigg]
\eea 
We have not managed to perform the sums needed to rewrite the action of $D^{(1)}$ on Gauss graph operators.
It is however straight forward to study this problem numerically, for specific choices of $n_{11}$ and $p$.

The numerical study we will discuss is focused on operators labeled by Young diagrams $R$ that have a 
total of $p=3$ long rows, and $n_{11}=3$.
The results of this example are rather typical.
A total of 21 operators can be defined, so that the dilatation operator is a $21\times 21$ dimensional matrix.
Acting on this space, $D^{(0)}$ decomposes into a block diagonal matrix with a total of 10 blocks.
Each block can be labeled by the vector $\vec{n}_{11}$. The possible blocks together with their dimension and 
allowed $s$ labels are
\bea
&&\vec{n}_{11}=(1,1,1)\qquad d=6\qquad s={\tiny \yng(3)\quad \yng(2,1)\quad \yng(1,1,1)}\cr
&&\vec{n}_{11}=(2,1,0)\qquad d=2\qquad s={\tiny \yng(3)\quad \yng(2,1)}\cr
&&\vec{n}_{11}=(2,0,1)\qquad d=2\qquad s={\tiny \yng(3)\quad \yng(2,1)}\cr
&&\vec{n}_{11}=(0,2,1)\qquad d=2\qquad s={\tiny \yng(3)\quad \yng(2,1)}\cr
&&\vec{n}_{11}=(1,2,0)\qquad d=2\qquad s={\tiny \yng(3)\quad \yng(2,1)}\cr
&&\vec{n}_{11}=(0,1,2)\qquad d=2\qquad s={\tiny \yng(3)\quad \yng(2,1)}\cr
&&\vec{n}_{11}=(1,0,2)\qquad d=2\qquad s={\tiny \yng(3)\quad \yng(2,1)}\cr
&&\vec{n}_{11}=(3,0,0)\qquad d=1\qquad s={\tiny \yng(3)}\cr
&&\vec{n}_{11}=(0,3,0)\qquad d=1\qquad s={\tiny \yng(3)}\cr
&&\vec{n}_{11}=(0,0,3)\qquad d=1\qquad s={\tiny \yng(3)}
\eea
It is a simple exercise to write down the complete set of partialy labeled Young diagrams\cite{Koch:2011hb} and write
down the action of the symmetric group on these states. 
We need to explicitely consider all 3 $\phi_{11}$-boxes as well as a single $\phi_{12}$ box when constructing the dilatation
operator numerically.
Within this space, the projectors $p_{r_{11}\gamma\beta}$ are $81\times 81$ dimensional matrices.
The only representation that carries a nontrivial multiplicity label is the $s={\tiny \yng(2,1)}$ representation in the
$\vec{n}_{11}=(1,1,1)$ subpace. 
The multiplicity free projectors can immediately be written down as
\bea
   p_{r_{11}\,\vec{n}_{11}}={d_{r_{11}}\over 3!}\sum_{\sigma\in S_3}\chi_{r_{11}}(\sigma)\Gamma^{\vec{n}_{11}}(\sigma)
\eea
with $\chi_{r_{11}}(\sigma)$ an $S_3$ character.
The matrix $\Gamma^{\vec{n}_{11}}(\sigma)$ represent $\sigma\in S_3$, in the displaced corners approximation and inside
the $\vec{n}_{11}$ subspace.
To construct the projectors for the $s={\tiny \yng(2,1)}$ representation in the $\vec{n}_{11}=(1,1,1)$ subpace, we need
to resolve this subspace into two $U(3)$ states in the ${\tiny \yng(2,1)}$ representation. 
The two states are described by the Gelfand-Tsetlin patterns that have the same inner multiplicity.
For our problem here, the two states are
\bea
\left[ 
\begin {array}{ccccc} 
    2  &            &    1    &           &      0    \\\noalign{\medskip}
       &     1      &         &     1     &           \\\noalign{\medskip}
       &            &    1    &           &      
\end {array} \right]\qquad\qquad
\left[ 
\begin {array}{ccccc} 
    2  &            &    1    &           &      0    \\\noalign{\medskip}
       &     2      &         &     0     &           \\\noalign{\medskip}
       &            &    1    &           &      
\end {array} \right]
\eea
and are easily constructed using $U(3)$ Clebsch-Gordan coefficients.
The detailed computation appears in Appendix C of \cite{Koch:2011hb}.

We find that $D^{(1)}$ not diagonal in the Gauss graph basis and it does not commute with $D^{(0)}$.
Further, it does not reduce to a block diagonal matrix and indeed, it mixes operators from different $\vec{n}_{11}$ sectors.
This mixing is expected and has a natural interpretation in the gravity dual.
Specifying $\vec{n}_{11}$ specifies how many oriented edges start and terminate at each node.
Interpreting the nodes as giant gravitons and the oriented edges as open strings attached to the giant graviton system,
$\vec{n}_{11}$ can only change as a result of open string splitting and joining.
Thus, the mixing we see is a signal of open string splitting and joining.
This interpretation is also natural given the fact that $D^{(1)}$ is a correction to the large $N$ limit, so that
we should indeed be seeing the first effects of string splitting and joining when this correction is included.
Finally, a remarkable feature of $D^{(0)}$ is the appearance of the integers $n_{ij}(\sigma)$ when the diagonalization
problem is solved.
Numerically we find that the eigenvalues of $D^{(1)}$ are again integers suggesting there may be a nice combinatorial
description of the problem, presumably exploiting the combinatorics of string splitting and joining.

\section{Discussion}\label{discussion}

In the $SU(2)$ sector of the ABJM theory we have managed to diagonalize the two loop dilatation operator
by employing the double coset ansatz.
This problem was already considered in \cite{deMelloKoch:2012kv} where the dilatation operator was already evaluated,
but not diagonalized.
One of the results we have reported, is precisely the solution of this diagonalization problem.
The main progress achieved in this article follows from our rewriting of the dilatation operator, in terms of adjoint variables.
This gives a useful organization of the dilatation operator and in particular, has allowed us to cleanly identify two terms that
contribute at the leading order at large $N$ and two that are subleading.
With this organization in hand, the eigenproblem of the dilatation operator is a straight forward exercise that can be achieved
using existing techniques.
The leading terms are diagonalized bythe  double coset ansatz, reducing the problem to the diagonalization of a collection of
decoupled oscillators, which is an integrable system.
We find a new ``conservation law'': the dilation operator does not mix operators with different $\vec{n}_{11}$ quantum number.
The resulting spectrum of anomalous dimensions differs from the corresponding spectrum in ${\cal N}=4$ 
super Yang-Mills theory in an important quantitative way.
In the ${\cal N}=4$ super Yang-Mills theory, the frequencies of the decoupled oscillators are set by the eigenvalues of
the matrix $n_{ij}(\sigma)$ which can be read straight from the permutation labeling the Gauss graph.
From (\ref{fdil}) we see that for ABJM the frequencies of the decoupled oscillators are set by the eigenvalues of
$(1+{r_{12,i}\over N})n_{ij}(\sigma)$. 
Thus, the frequencies depend both on the matrix $n_{ij}(\sigma)$, determined by the Gauss graph, and on $r_{12,i}$
which are the row lengths of the Young diagram $r_{12}$.
Each row of $r_{12}$ corresponds to a giant graviton.
The number of boxes in the $i^{\rm th}$ row of $r_{12}$ determines an ${\cal R}$- charge which corresponds to the
angular momentum of the giant graviton.
Since the giant expands to a definite size by balancing a Lorentz type force (trying to expand the giant) with tension (trying to
shrink the giant), the angular momentum of the giant sets the size of the giant.
Consequently, our result implies that the excitation spectrum of the giant graviton picks up a dependence on the size 
of the giant graviton.
The fact that the spectrum of the anomalous dimensions in ${\cal N}=4$ super Yang-Mills theory is independent of the
parameters of the Young diagram associated to the giant graviton system, matches the fact that the spectrum of small
fluctuations around the giant is independent of the size of the giant\cite{Das:2000fu}.
This independence of the size of the giant is understood as follows\cite{Balasubramanian:2004nb}: as the radius of the giant
increases, there is an increase in the energy of fluctuations due to blue-shifting, as well as a decrease in the energy of the states
because the fluctuations now move on a bigger sphere.
These two effects precisely cancel producing a size independent spectrum.
For the ABJM case, our results predict that although these two effects still operate, they do not precisely cancel so that
the spectrum does pick up a dependence on the size of the giant. 
This is consistent with the small fluctuation spectrum around a giant graviton performed in \cite{Giovannoni:2011pn}.
By perturbing around the near-maximal giant and the ``small" giant these authors find a spectrum that is size-dependent. 

In this article we have also given a simple formula for the normalization of the Gauss Graph operators.
This will be a useful technical input when computing the effects of Gauss Graph operator mixing, at subleading orders in a
large $N$ expansion.

Finally, we have also evaluated the largest of the subleading (in ${1\over N}$) terms.
Although we have not managed an analytic result, a numerical study has lead to some interesting conclusions.
The subleading correction does not commute with the leading order dilatation operator.
Further, it allows mixing between operators with different $\vec{n}_{11}$ quantum numbers, so that
it spoils the conservation law that was present at large $N$.
This is naturally interpreted as a consequence of open string splitting and joining.
The discusion of \cite{Kristjansen:2008ib,Caputa:2009ug} suggests that the failure of this conservation law may
be an indication that integrability does not persist beyond the large $N$ limit.
A numerical diagonalization of this term shows that it has integer eigenvalues, suggesting that there may be a nice 
combinatorial description waiting to be developed.

{\vskip 0.2cm}

\noindent
{\it Acknowledgements:}
We thank Jeff Murugan for helpful correspondence.
 RdMK and RK are supported by the South African Research Chairs
Initiative of the Department of Science and Technology and the National Research Foundation.
SS is supported by the National Institute for Theoretical Physics.

\begin{appendix}

\section{Normalization of the Gauss Graph Operators}

The two point function of Gauss Graph Operators is
\bea
   \langle O_{R,r}(\sigma)^\dagger  O_{R,r}(\sigma)\rangle =\sum_{\gamma_1,\gamma_2\in H}
                          \delta (\sigma^{-1}\gamma_1\sigma\gamma_2^{-1})
\eea
The right hand side of the above equation is simply counting the number of solutions $\gamma_1,\gamma_2\in H$ to
\bea
    \sigma = \gamma_1\sigma\gamma_2^{-1}\label{sc}
\eea
Using $\gamma_1$ and $\gamma_2$ we are able to swap the endpoints of the open strings. 
If we swap the labels of strings that have the same start and endpoints, we leave $\sigma$ unchanged
and hence have a solution to (\ref{sc}).
In this way, for $n$ strings stretching from the same start point to the same endpoint, we will pick up a factor of $n!$.
Denote the number of oriented line segments stretching from node $i$ to node $j$ by $n_{ij}$ and the number of 
segments stretching from node $i$ back to node $i$ by $n_{ii}$.
We have
\bea
   \langle O_{R,r}(\sigma)^\dagger  O_{R,r}(\sigma)\rangle =\prod_{i=1}^p n_{ii}!\prod_{k,l=1,l\ne k}^p n_{kl}!
\eea

\end{appendix}

\end{document}